\documentclass[12pt]{iopart}

\usepackage{iopams}  
 \expandafter\let\csname equation*\endcsname\relax
  \expandafter\let\csname endequation*\endcsname\relax
\usepackage{graphicx,amsmath,amssymb}
\usepackage{verbatim}
\usepackage{ragged2e}
\usepackage{color}
\usepackage{braket}
\usepackage{changes}

\newcommand{\LN}{LiNbO$_3$}

\newcommand{\Fig}[1]{Fig.~\ref{#1}}

\newcommand{\TE}{_{\mbox{\small{TE}}}}
\newcommand{\TM}{_{\mbox{\small{TM}}}}

\begin{document}

\title{Toolbox for the design of \LN-based passive and active integrated quantum circuits}

\author{P. R. Sharapova$^{1}$, K. H. Luo$^{1}$, H. Herrmann$^{1}$, M. Reichelt$^{1}$,
T.~Meier$^{1}$ and C. Silberhorn$^{1}$}

\address{$^1$Department of Physics and CeOPP, University of Paderborn,
Warburger Strasse 100, D-33098 Paderborn, Germany}
\ead{polina.sharapova@upb.de}
\vspace{10pt}
\begin{indented}
\item[]\today
\end{indented}

\begin{abstract}
We present and discuss perspectives of current  developments on advanced quantum optical circuits monolithically integrated in the lithium niobate platform. A set of basic components comprising photon pair sources based on parametric down conversion (PDC), passive routing elements and active electro-optically controllable switches and polarisation converters are building blocks of a toolbox which is the basis for a broad range of diverse quantum circuits. We review the state-of-the-art of these components and provide models that properly describe their performance in quantum circuits. As an example for applications of these models we discuss design issues for a circuit providing on-chip two-photon interference.  The circuit comprises a PDC section for photon pair generation followed by an actively controllable modified Mach-Zehnder structure for observing Hong-Ou-Mandel (HOM) interference. The performance of such a chip is simulated theoretically by taking even imperfections of the properties of the individual components into account. 
\end{abstract}

%
%
%
%
%

\section{Introduction}

Recent research on quantum communication and information processing (QCIP)  has proven that quantum mechanical effects
can be applied to overcome classical performance limitations in the fields of communication security, metrology and computation.
Future practical development of QCIP will strongly rely on the development of miniaturised and reliable optical components.
Over the past several years complex integrated photonic circuits have already been demonstrated, which, for
example, allow for the realisation of on-chip two-photon interference, CNOT-gates and path-entangled
states of two photons \cite{PolitiScience2008}, multi-photon quantum interference  \cite{PeruzzoNC2011}, the simulation of quantum Boson sampling \cite{SpringScience2012} and quantum walk systems \cite{PeruzzoScience2010,CrespiNP2013}. These works have demonstrated the benefits
of integrated solutions \cite{Tanzilli_Genesis,OBrienNJP2013} over bulk optical approaches in terms of stability, scalability, connectability
and reproducibility.  In addition, due to its on-chip nature and controllable capability, the miniaturised optical system represents a significant step forward for achieving integrated all-optical communication and optical computing systems.

However, the development of compact quantum optical circuits using waveguide components is still at a very early stage. The essential building blocks of such circuits are photon sources, passive and active elements to route and manipulate the photons and  -- in the ideal case -- even single photon detectors. All these components with good performance properties  have been demonstrated as single devices, but were realised in different materials. Thus, nowadays it is still an open question which material platform is best suited for the integration of larger advanced quantum circuits~\cite{Bogdanov}. 

\LN\ is an attractive candidate serving as a platform for the development of advanced integrated quantum circuits. Besides its outstanding electro-optic, acousto-optic   and nonlinear optical properties  \cite{Gaylord-Weiss, KKWong}, mature and reliable waveguide fabrication techniques, e.g., by the indiffusion of Titanium, exist, which is a basic prerequisite. On-chip generation and manipulation of qubits can harness the strong $\chi^{(2)}$-nonlinearity which can be specifically tailored by exploiting quasi-phase-matching (QPM) in periodically-poled waveguides \cite{HumCRP2007}. For instance, photon pairs can be generated efficiently via parametric down-conversion (PDC). The exploitation of the versatile electro-optic properties of \LN\ enables the manipulation of qubits (e.g. phase, polarisation, etc.) with bandwidths up to several GHz. However, the full potential of \LN\ has not yet been completely exploited. Various groups have demonstrated the photon-pair generation using type I or type II phase-matching in different configurations and different wavelength ranges~\cite{
TanzilliEPJD2002,Jiang_OptCom2006,Suhara_PTL2006,Suhara_PTL_2007,Fuji,
Martin_TypeII_PDC,Herrmann_OptExp2013, Lim, Zhang, Bock, Martin, Halder, Takesue, Jiang}, but little work has been done in establishing a framework, which combines several non-linear and active devices in particular for including electro-optic functionality in more complex structures.

Towards a more complex integration a few examples have recently been demonstrated; reviews can be found in \cite{Tanzilli_Genesis,Alibart_Quantum_Photonics}.  For example in \cite{Krapick} an efficient source of heralded  single photons using non-degenerate PDC followed by a passive directional coupler acting as wavelength selective splitter has been demonstrated. More advanced quantum circuits are a quantum relay chip~\cite{Nice_Relay} and a circuit for the on-chip generation and manipulation of entangled photons \cite{Jin}, both of these examples already include active electrooptic components \cite{Bonneau, Lenzini}.

In this paper, we discuss design issues  of advanced integrated quantum circuits in \LN  with a focus on multiple functionalities in a multi-channel structures, which include passive components as well as sources and active electrooptic devices. We start with a review of the operational principles and the state-of-the-art of the elementary building blocks which are required for such circuits. For each of these components  we present a model for a quantum optical description which allows also to take into account possible imperfections and phase-matching conditions.  As an example we discuss the possibility of an on-chip two-photon interference circuit comprising a PDC stage for photon pair generation followed by an actively controllable modified Mach-Zehnder structure for Hong-Ou-Mandel (HOM) interference.  

\section{Functional Elements for Quantum Circuits}
In the past decades,  a multitude of integrated optical devices in \LN\ have been developed and demonstrated \cite{Alferness,Sohler_OPN}. In particular, electro-optic modulators  have become successful commercial products, which are dedicated for classical applications in telecommunications. However, all of these devices are typically implemented as separate standalone devices and are not combined in a single multi-channel structure with multiple functionalities. Our vision is to benefit from these existing experiences and adapt them to the challenging demands required for existing quantum optical applications, which rely on several distinct components on a single chip. A prerequisite for the development of advanced integrated quantum circuits in \LN\ is a profound understanding and optimisation of the individual components forming such circuits.

The basis for the development of such devices is a reliable waveguide technology. Two mature waveguide fabrication techniques exists for integrated optical devices in \LN: proton-exchange  and Ti-indiffusion \cite{Alferness}. Both techniques enable the fabrication of single-mode waveguides with low losses ($< 0.1~$dB/cm). However, only the Ti-indiffusion technique provides guiding of waves in both orthogonal polarisations (TE and TM or H and V). 

Besides these renowned waveguide fabrication techniques, in the past two decades ridge type waveguides\cite{Nishida,Kishimoto,Umeki} in (periodically poled) \LN\ have also attracted a lot of attention because such ridge guides show less sensitivity to photo-refraction and, thus, can be operated at higher optical power levels without degradation due to ``optical damage".  In a first process step, planar waveguides are fabricated either by wafer bonding (doped) \LN\ on undoped LN \cite{Nishida} or LiTaO$_3$ \cite{Umeki}  or by a planar proton exchange \cite{Kishimoto}. Subsequently, ridges are formed either by diamond blade dicing \cite{Nishida,Kishimoto} or dry etching \cite{Umeki}. Although the performance of nonlinear optical devices achieved with such ridge waveguides is impressive, there are several drawbacks for implementing them into advanced circuits: Ridge structures fabricated by dicing techniques are limited to straight guides, i.e.\ the implementation of curved bendings, which is definitely necessary for complex circuits, is hardly compatible with this fabrication method. Moreover, the non-planar surface after waveguide fabrication severely complicates subsequent fabrication steps like, for instance, the deposition of electrodes for electro-optic devices.

For the reasons mentioned above, we believe that the Ti-indiffusion technique is the preferred one, in particular, if polarisation qubits are exploited which require waveguiding for both polarisations. Therefore,  we restrict the discussion to devices compatible with the Ti-indiffusion technology. In the following we will discuss the design of different functional blocks to build up integrated quantum photonic devices forming a “toolbox” for application specific quantum circuits. In order to estimate the feasibility of combining different components to a complex circuitry in a practical  system, we present the classical characterisation of theses elements which serve as reference points for the evaluation of needed accuracies. The “toolbox” contains elements to generate and convert single photons or photon pairs via $\chi^{(2)}$-based nonlinear interactions, truly passive elements to route photons on-chip and electro-optic actively controllable elements for switching and polarisation conversion.  In the following we discuss in detail the specific features of these different categories.

\subsection{Passive Routing}

\begin{figure}[htbp]
\begin{center}
\includegraphics[width=6cm]{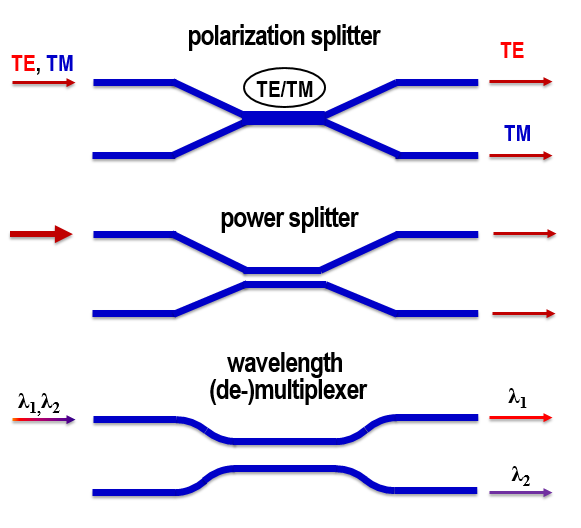}
\end{center}
 \caption{\label{Passive_Box} Basic structure of the directional coupler with  zero gap and large gap acting as integrated polarisation and/or power splitter and wavelength multiplexer.}
\end{figure}

We start our description of the toolbox for  the \LN  circuits with passive routing devices. The operating principle of these components is universal for all integrated platforms, but the specific experimental parameters of the materials define the specific guiding properties  of TE and TM polarisations, which in turn define achievable performance parameters of an overall structure. Directional couplers are the basic elements for the routing and splitting of the biphoton states. Such structures comprise a coupling region consisting of two closely adjacent waveguides in which a power exchange between these guides is possible. A proper design of this coupling region enables specific functionalities of the directional couplers. In \Fig{Passive_Box} some examples are sketched.

To achieve the desired functionality the specific geometries of a distinct structure has to be determined. This includes the size of the waveguides, the separation of the waveguides in the coupling region, the length of the coupling region and the coupling in the branching regions. Several modelling tools for such directional couplers are available.

In the local normal mode description~\cite{Bersiner}, the overall coupler is treated as a single waveguide with changing dimensions along the propagation direction. At least two eigenmodes - a symmetric and an antisymmetric - are guided in this structure. Launching a wave into one input port of the directional coupler will excite the two eigenmodes with equal amplitude. Assuming that all dimensional changes along the coupler structure are adiabatic, there is no coupling, i.e. power exchange, between these eigenmodes. But the propagation constants of the eigenmodes along the structure may vary.  The power splitting at the output of the coupler is a result of the interference of the two eigenmodes at the end of the structure. If the accumulated phase-difference between the even and odd eigenmodes is an even multiple of $\pi$, the power is routed to the bar-state output. For an odd multiple of $\pi$ the power is coupled into the cross-state output.

\begin{figure}[h]
\begin{center}
\includegraphics[width=7.5cm]{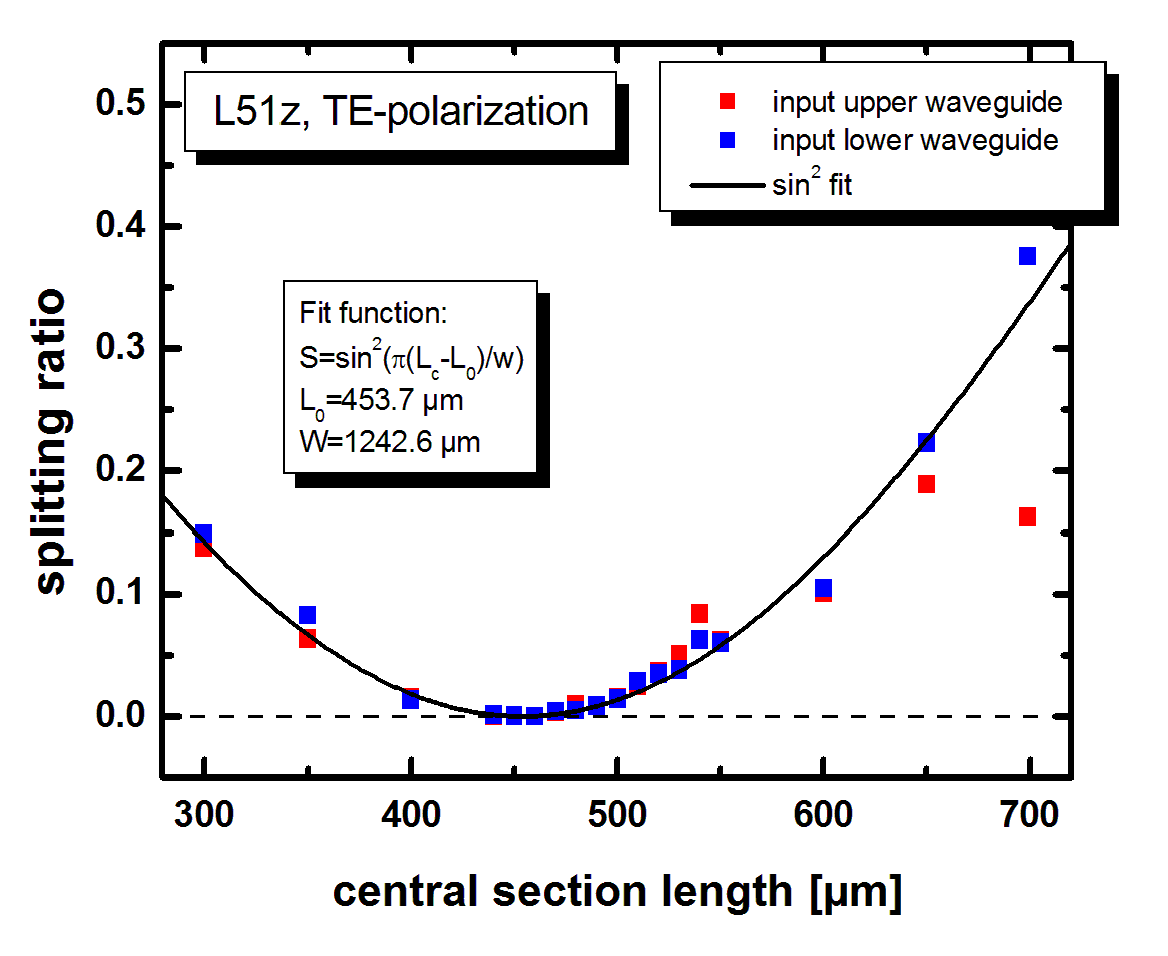}
\includegraphics[width=7.5cm]{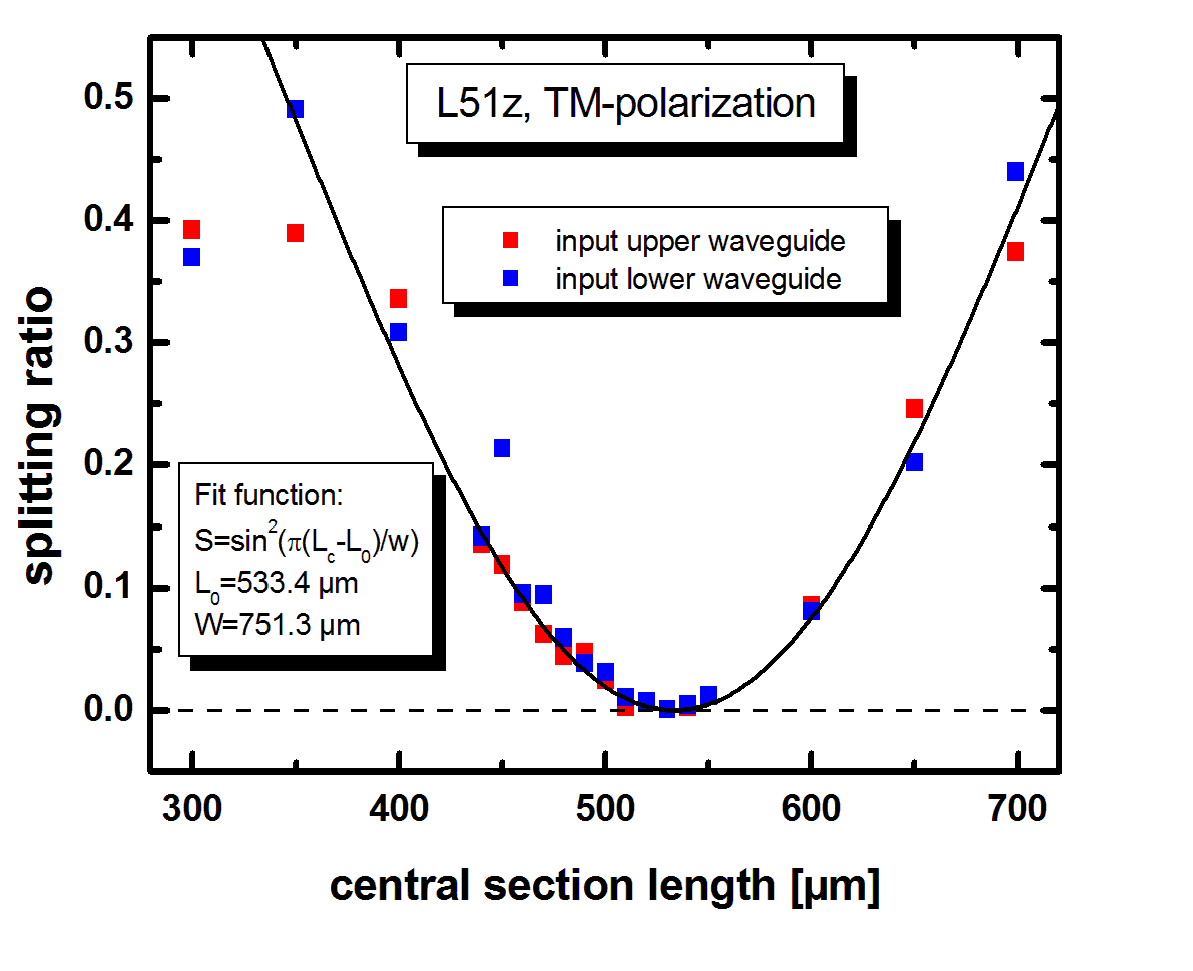}
\end{center}
 \caption{\label{L51z_Splitting} Measured splitting ratios of a typical sample as function  of the central section length by using the narrow band laser $\lambda_p=1550$ nm. Additionally, a sin$^2$-fit function to the data is shown. Please note that the experimental data are shown with error bars smaller than symbol size.}
\end{figure}

In \LN\ various types of directional couplers for different applications have been studied (see e.g. \cite{Alferness}). An example of the splitting performance of a directional coupler, which was designed to operate as a polarisation splitter is shown in \Fig{L51z_Splitting}.   The coupler consists of a central section, in which the two incoming waveguides are merged to a single one with doubled width, i.e.\ there is no gap between the waveguides anymore (zero-gap coupler). The separation of the waveguides in the branching regions is done via inclined straight waveguides.  The branching angle and the length of the central coupling region was designed using the local normal mode method described above. The accumulated phase-difference of the eigenmodes is $2\pi$ for TE-polarised waves and $3\pi$ for TM-polarisation. Thus, TE waves are routed to the bar-state output and TM-waves to the cross-state output.

In order obtain realistic estimates for  the manufacturability  of experimental of \LN circuits with realistic margins we fabricated such structures with a certain variation of the central section length. The splitting ratio is defined as the ratio of the power in the unwanted port to the sum of the powers in both output ports. In \Fig{L51z_Splitting} the measured splitting ratios for TE and TM are shown. It can be seen that the device really works as polarisation splitter if the central section length is about 500 $\mu$m. But the best splitting ratios for TE and TM occur at slightly different central section lengths. The optimum for TE is achieved with $L_c\approx450~ \mu$m whereas the ideal coupling for TM is at $L_c\approx 530~\mu$m. However, even with $L_c\approx500~\mu$m a polarisation splitting better than 17 dB for both polarisations is obtained.

To describe the propagation of a biphoton state through such a polarisation splitter a unitary transformation can be used.
We can describe the corresponding transformation matrix of the PBS element in the basis
$\{a_{1H}^{\dagger}(\omega_{s,i}),a_{1V}^{\dagger}(\omega_{s,i}),a_{2H}^{\dagger}(\omega_{s,i}),a_{2V}^{\dagger}(\omega_{s,i})\}$,
where the index $\{1,2\}$ denotes the channel number and
$\{H,V\}$ the polarisation, by
\begin{equation}
PBS=\begin{pmatrix}
 i \sin \alpha & 0 & \cos \alpha & 0 \\
  0 & i \cos \beta & 0 & \sin \beta \\
 \cos \alpha & 0 & i \sin \alpha & 0 \\
 0 & \sin \beta & 0 & i \cos \beta\\
 \end{pmatrix}.
\label{PBS}
\end{equation}

 Here, the angle $\alpha$ describes the transformation of the
horizontally-polarised photons and the angle $\beta$ corresponds to the vertically-polarised photons.
The PBS element works perfect  for $\alpha=\beta =\pi/2$ and in this case the
horizontally-polarised mode is transmitted in the upper channel 1 and the vertically-polarised mode in the lower channel 2, respectively.
The modelling of the PBS with the two angles $\alpha$ and $\beta$ as free parameters allows us to describe different imperfect polarisation splittings for TE and TM. As can be seen from \Fig{L51z_Splitting} the splitting ratios vary only slowly with wavelength. Thus, any wavelength dependence of the operation of such splitters  on the biphoton wavefunctions can be neglected as long as the frequencies of the two photons are close to degeneracy.

As a second passive element we consider a beam splitter (BS) and wavelength multiplexer, where we also need to take into account fabrication parameters. The beam splitter for power splitting or wavelength multiplexing is usually obtained by using a directional coupler with a large gap and  S-shaped bends consisting of circular arcs  \cite{Alferness}. The  central gap separation, bending radius and  central section length has to be  optimised to provide the desired purpose. As soon as the coupler is fabricated with the designed parameters, its splitting behaviour is fixed passively. It should be noted that the performance of such couplers with large gap are very sensitive to fabrication tolerances \cite{Alferness_Directional_Couplers} and also their wavelength dependence is more critical compared to zero-gap couplers \cite{Januar}. 

For taking imperfections in the different channels of the BS into account  we introduce two angles $\theta$ (for the upper channel) and $\xi$ (for the lower channel) and define the common BS transformation by
\begin{equation}
BS=\begin{pmatrix}
  \cos \theta & 0 & i \sin \theta  & 0 \\
   0 &  \cos \xi & 0 & i\sin \xi \\
i\sin \theta & 0 & \cos \theta & 0 \\
 0 & i \sin \xi & 0 & \cos \xi\\
 \end{pmatrix} .
\label{BS}
\end{equation}

\subsection{Photon Pair Sources}

\begin{figure}[htbp]
\begin{center}
\includegraphics[width=6cm]{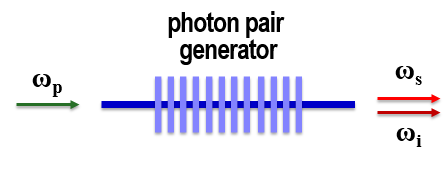}
\end{center}
 \caption{\label{Generation_Box} Photon pair generation by parametric down-conversion (PDC) in a periodically-poled \LN\ waveguide. }
\end{figure}

As a second important ingredient \LN quantum circuits offer the possibility to include highly efficient parametric down-conversion sources for the generation of photon pairs. The development of telecom compatible, compact and reliable sources of entangled photons is a prerequisite to pave the way  for most quantum optical experiments, either for fundamental research or even for a successful implementation of future quantum information technologies into real world applications. Therefore, the development of rugged and efficient photon (pair) sources to produce qubits have been developed over the last decades in various contexts. Photon pairs  can be generated via PDC exploiting the $\chi^{(2)}$ nonlinearity of LiNbO$_3$ in a waveguide channel, such that they can be directly incorporated in complex circuitries with different functionalities  (\Fig{Generation_Box}). 


\begin{figure}[htbp]
\begin{center}
\includegraphics[width=14cm]{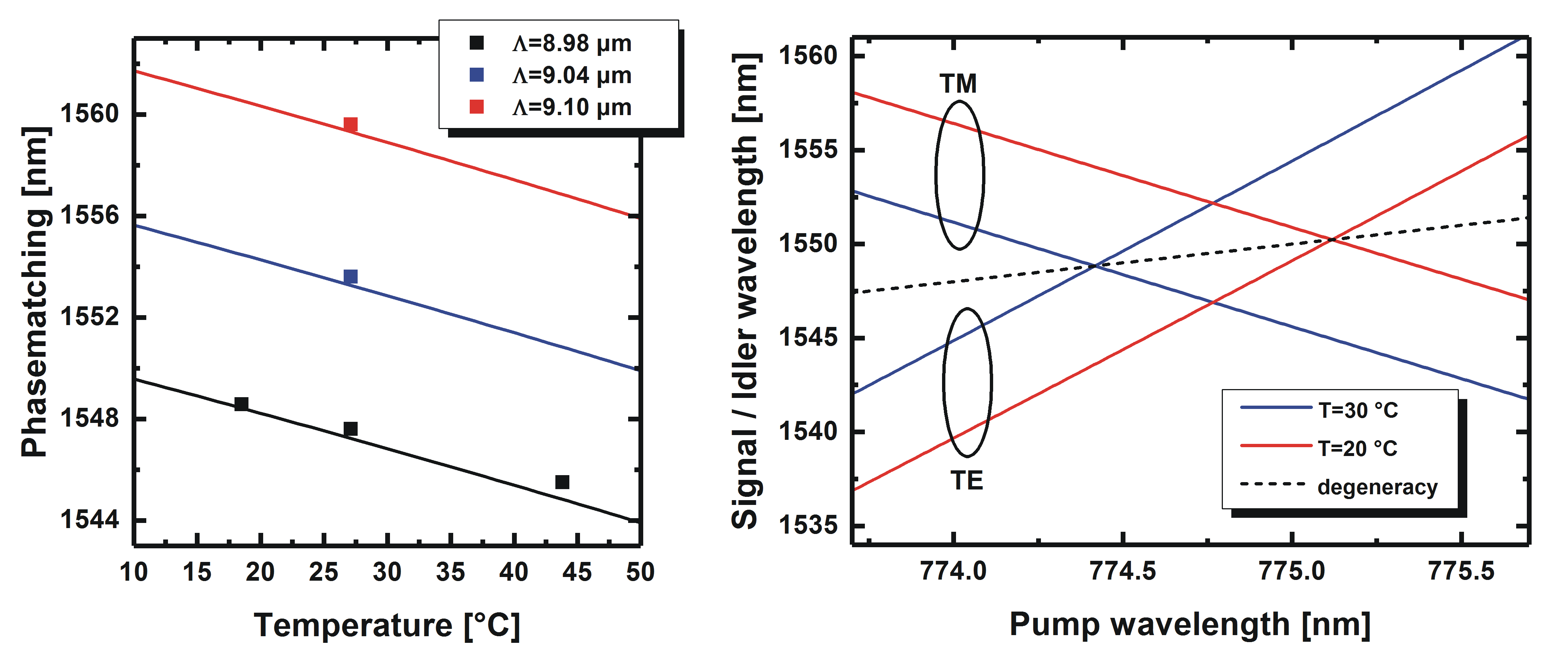}
\end{center}
 \caption{\label{PM_SHG} Type II phasematching for PDC. Left: signal/idler wavelength for degenerate PDC in a Ti-indiffused waveguide in PPLN for various poling periods as function of the sample temperature. Solid lines: Calculated curves based on Sellmeier equations for LiNbO$_3$ \cite{EdwardsLawrence,Jundt} and a waveguide dispersion derived using a Gauss-Hermite-Gauss approximation \cite{GHG}. Symbols: Experimental data determined via second harmonic generation (SHG). Right:
Calculated signal and idler wavelengths for phase-matched PDC as function of the pump wavelength assuming a fixed poling period of $\Lambda=9.0~\mu$m. Calculated curves for $T=20 ~^o$C  (red) and $T=30 ~^o$C (blue) are shown. The dashed curve marks degeneracy.}
 
\end{figure}



The photon pair generation via PDC is governed by energy conservation and phasematching. The latter one requires a matching of the wave vector of the pump photon $k_p$ with the wave vectors of the generated signal and idler photon $k_{s,i}$  according to:
\begin{equation}
k_p=k_s+k_i-\frac{2\pi}{\Lambda},
\end{equation}
where $\Lambda$ is the period of the ferroelectric domain grating in the periodically-poled \LN\ waveguide (PPLN). Via a proper choice of this poling period $\Lambda$ the wavelengths at which phase-matching is fulfilled can be determined. Of course, $\Lambda$ is fixed during the fabrication process. However, a fine tuning to the desired wavelength combination is usually possible by exploiting  the temperature dependencies of the refractive indices.

Different phase-matched PDC processes can be exploited: Via a type I phase-matching, extraordinarily polarised signal and idler photons are created by an extraordinarily (TM/H) polarised pump photon.  This process uses the strong $d_{33}$ nonlinear coefficient of \LN\ and, thus, is very effective. However, for a process close to degeneracy, the spectral bandwidth is very broad
(typically some 10 nm). Another option is a type II phase-matched process, where an ordinarily polarised pump generates a photon pair with orthogonal polarisation via a nonlinear coupling process which exploits the $d_{31}$ coefficient. The efficiency of this process is about $\left(d_{33}/d_{31}\right)^2\approx 40$ times smaller, but the spectral bandwidth is also much narrower. Thus, the spectral densities of both processes are almost equal~\cite{Fuji}. As a source of qubits preparation the type II phase-matched degenerate PDC is in most cases preferable for implementing quantum circuits on a single platform as it generates  signal and idler photons with orthogonal polarisations but equal frequencies which can be directly used to encode the information and to further process it in linear networks.

As an example of a typical PDC source, which is compatible with quantum circuit designs, we analyse the experimental data for such a source in dependence of the relevant parameters. For a degenerate type II phase-matched PDC process to generate photon pairs in the telecom band poling periods around 9 $\mu$m are required. Details are shown in \Fig{PM_SHG}. The left diagram shows the phase-matching wavelength for degenerate PDC versus the sample temperature for various poling periods. Experimental data have been obtained via second harmonic generation (SHG), which is the reverse process to degenerate PDC. The measured data correspond very well with the calculated ones, which were obtained using Sellmeier equations for bulk \LN\ \cite{EdwardsLawrence,Jundt} and by determining the waveguide dispersion via a simple Gauss-Hermite-Gauss approximation as described in \cite{GHG}. For a fixed poling period the temperature dependence of the degeneracy wavelength is given by $\approx -0.14~$nm/$^o$C.

On the right hand side of \Fig{PM_SHG} the phase-matching behaviour as function of the wavelength of the pump, which is assumed to be a monochromatic wave, is shown. The diagram shows signal and idler versus pump wavelength, which were calculated for a fixed poling period of $\Lambda=9.0~\mu$m and two different temperatures. The cross-like structure of the phase-matching curves leads to a separation of the peak wavelengths for signal and idler, if the pump wavelength not exactly matches the condition to obtain degenerate PDC. In the presence of a deviation $\delta\lambda_p$ from that degeneracy point, the difference between signal and idler wavelength is about $|\lambda_s-\lambda_i|\approx 15\cdot\delta\lambda_p$.

For a type II degenerate PDC in a periodically-poled \LN\ section with waveguiding achieved by Ti-indiffusion, the signal and idler photons of the photon pair have orthogonal polarisations, they have different group velocities in the waveguides.
The biphoton state which is created in the PDC section can be described according to lowest-order perturbation theory by the following state vector \cite{Law}
\begin{equation}
\Ket{\psi_{PDC}}=\int d \omega_s d \omega_i F(\omega_s, \omega_i)a^\dagger_{1H}(\omega_s)a^\dagger_{1V}(\omega_i)\Ket{0},
\label{input}
\end{equation}
where $a^\dagger_{1H}(\omega_s)$ and $a^\dagger_{1V}(\omega_i)$  are the photon creation operators for the signal (idler) frequency modes which are horizontally and vertically polarised, respectively, and the index $1$ corresponds to a particular spatial mode, later referred to as channel, in the waveguide system.
The two-photon amplitude (TPA) $F$ represents the probability amplitude of the creation of photons with frequencies $\omega_s$ and $\omega_i$.  For the PDC section pumped by a laser with a Gaussian envelope, $E_p^{(+)} (\mathbf{r},t)= E_0 e^{-\frac{t^2}{2\tau^2}}e^{i( \mathbf{k}_p \cdot \mathbf{r}-\omega_p t)}$,  with a full width at half maximum (FWHM) of the intensity of $2\sqrt{\ln 2}\tau$, $\tau$ is a pulse duration, the TPA is given by
\begin{eqnarray}
F(\omega_s, \omega_i)=C \exp [\frac{(\omega_s+\omega_i-\omega_p)^2}{2\Omega^2}]\mathrm{Sinc}[\frac{\Delta k \, L_{PDC}}{2}]
\exp[\frac{i\Delta k\,  L_{PDC}}{2}], \ \ \ \
\label{TPA}
\end{eqnarray}
where $C$ is the normalisation constant, $\Delta k =k_p(\omega_s+\omega_i)-k_s(\omega_s)-k_i(\omega_i)+\frac{2 \pi}{\Lambda}$ is the wave vector mismatch inside the periodically-poled section, $\Lambda\approx 9~ \mu$m is the poling period, $L_{PDC}$ is the length of the PDC section
and $\Omega=1/ \tau$. The signal and idler wave vectors $k_{s,i}=\frac{n_{s,i}(\omega_{s,i})\omega_{s,i}}{c}$ depend on the effective indices $n_{s,i}(\omega_{s,i})$ of the waveguide modes  which can be calculated as $n_{s,i}=n_{s,i}^{LiNbO_3}+\delta n_{s,i}$. The first term is the bulk refractive index of \LN\ which can be obtained by using the dispersive Sellmeier formulas  \cite{Sellm}. To determine $\delta n_{s,i}$ the propagation constants of the respective waveguide modes must be calculated. They can be obtained using a numerical solver (e.g.\ a finite element based solver) applied to the refractive index profile of the Ti-indiffused waveguide  \cite{Strake_Bava_Montrosset}. In most cases, however, much simpler almost analytical methods (e.g. Gauss-Hermite-Gauss approximation~\cite{GHG} or effective-index-methods~\cite{Strake_Bava_Montrosset}), provide sufficiently good results for a quantitative evaluation.

The TPA is mainly characterised by two parameters: the duration of the pump pulse and the length of the PDC section. Note that whereas the brightness of the PDC source increases with the length of the PDC section, the increased length simultaneously gives rise to a time delay between the orthogonally polarised photons. This is caused by the fact that the PDC process predominantly takes place in the centre of the PDC section.
This effect is included in our description by the last term in Eq.~(\ref{TPA}).  Therefore, already at the end of the PDC section the two photons have accumulated a time delay determined by approximately their propagation with different velocities through half of the PDC section length. For embedding such PDC source in structures with different components the design of appropriate geometries has to take care of appropriate delay lines within the monolithic structures. This constraint can cause important restrictions for the feasibility of network designs. 

The efficiency of the generation process depends on the nonlinear optical coefficient, the overlap of the pump field mode with the signal and idler mode and on the interaction length. The overlap can be optimised by a proper design of the waveguide geometry and, hence, the waveguide fabrication. Detailed studies on optimised profiles have been performed for various classical nonlinear three wave mixing processes like second harmonic, sum- or difference frequency generation~\cite{Suhara_Book}.

Several papers reported the application of such a type II phase-matched PDC process in Ti-indiffused PPLN waveguides for photon pair generation in the telecom band \cite{Suhara_PTL2006, Martin_TypeII_PDC,Fuji}. In a 30 mm long waveguide efficient pair generation within a spectral bandwidth $<$1~nm and a brightness of $6\times10^5~$pairs/(s~mW~GHz) has been demonstrated~\cite{Fuji}. Such efficiencies are about two orders of magnitude larger compared to sources exploiting PDC in bulk crystals.

\subsection{Active Electrooptic Manipulation}

\begin{figure}[htbp]
\begin{center}
\includegraphics[width=6cm]{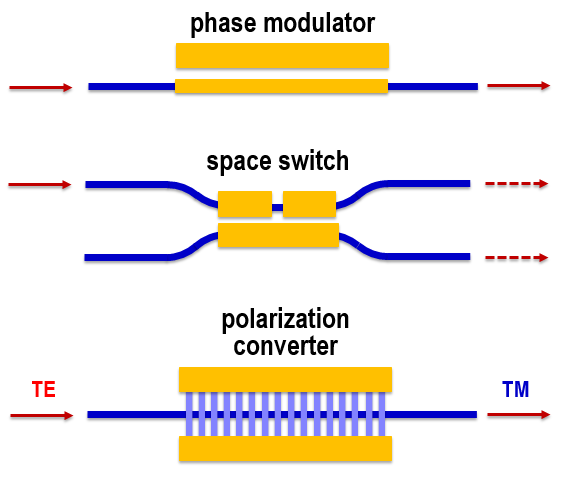}
\end{center}
 \caption{\label{Active_Box} Basic structures of the active manipulation in integrated waveguides.}
\end{figure}

An outstanding feature of \LN\ is its capability of electrooptic modulation which enables on-chip manipulation of optical waves via externally applied control voltages. These electrooptic properties have been widely used to develop various versions of e.g.\ integrated intensity modulators or phase-shifters \cite{Alferness,Wooten}. Obviously, the exploitation of electrooptic manipulation of optical waves offers also a huge potential for the design and realisation of advanced integrated quantum circuits. However, their inclusion into more complex circuit structures have not yet been addressed in the theoretical modelling of practical quantum networks. Examples of such electro-optic devices for the active manipulation of the quantum state are given schematically in  \Fig{Active_Box}.

\subsubsection{Phase modulators}

The simplest electro-optic device is a phase-modulator which comprises an optical waveguide within an electric field. This field is generated via a voltage applied to electrodes. For a phase-modulator in z-cut \LN\ one electrode is placed on top of the waveguide (but is separated from the waveguide with a dielectric buffer layer) and a second electrode besides the waveguide. A  voltage applied to the electrodes generates an electrical field which is predominantly oriented along the crystalline z-direction in the waveguide region.  The phase of a TM-polarised wave is modulated by the electric field via the strong electro-optic coefficient $r_{33}$ whereas TE-polarised waves are modulated via $r_{31}$. As the latter coefficient is about a factor of three smaller, the overall device performance yields a differential phase-modulation.

Due to the fact that the phase modulator changes the phase of the vertically and horizontally polarised modes differently, the matrix describing a phase modulator placed in one spatial channel is:
\begin{equation}
PM=\begin{pmatrix}
  \exp[i \Phi_H] & 0 & 0  & 0 \\
   0 & \exp[i \Phi_V] & 0 & 0 \\
0 & 0 &  1 & 0 \\
0 & 0& 0 & 1\\
 \end{pmatrix},
\label{PM}
\end{equation}
where $\Phi_H$ and $\Phi_V$ are the additional phase of the horizontally and vertically polarised modes, respectively.

\subsubsection{Electro-optically adjustable beam splitters}

Variable beam-splitters are key components for various quantum applications which require any kind of dynamic spatial routing or switching or solely a fine-tuning of the splitting ratios. For such demands, electro-optical beam splitters are best suited as they can provide a flexible operation even at high bandwidths, which in principle can go up to several GHz.

\begin{figure}[tbhp]
\begin{center}
\includegraphics[width=8cm]{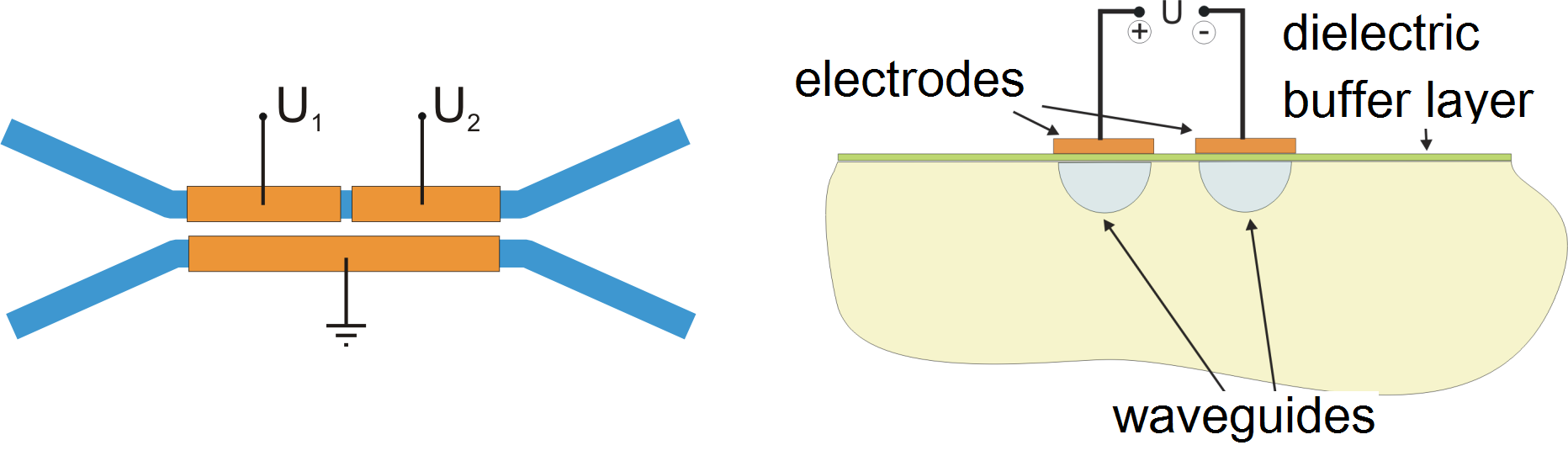}
\end{center}
\caption{\label{DeltaBetaCoupler}Schematics of an electro-optical switch comprising of a directional coupler with $\Delta\beta$-reversal  electrodes. Left: Top view, right: cross-sectional view in the coupling region.}
\end{figure}

The basis of such electro-optic splitters are again directional couplers which need to be equipped with proper electrodes for the electro-optic driving. The most flexible configuration is based on a so called $\Delta\beta$-reversal scheme as is sketched in \Fig{DeltaBetaCoupler} \cite{Kogelnik}. Usually, a structure with a relatively large gap between the waveguides in the coupling region is chosen. Thus, the weak coupling leads to a longer coupling length, which is required to provide a sufficient interaction length for the electro-optic modulation. The electrode configuration is sketched in \Fig{DeltaBetaCoupler}. Electrodes are above the two optical waveguides but separated via a dielectric buffer layer. Above one waveguide the electrode is split in two parts, which allows to apply two different voltages $U_1$ and $U_2$. Such an electrode scheme allows to compensate even if some inhomogeneities - like for instance a slight mismatch of the width of the two waveguides in the coupling section - are present. By a proper choice of the two voltages a perfect switching between the bar- and the cross-state output ports is possible.

\begin{figure}[htb]
\begin{center}
\includegraphics[width=6.8cm]{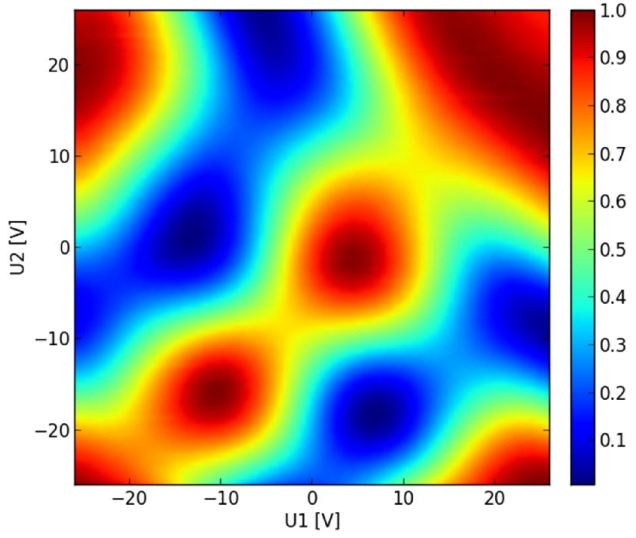}
\end{center}
\caption{\label{VoltageScan} Measured (normalised) bar state output power as function of the applied voltages $U_1$ and $U_2$ of an electro-optically adjustable beam-splitter consisting of a directional coupler with an 8 mm long coupling region. }
\end{figure}

In \Fig{VoltageScan} an experimentally determined switching characteristics of such a directional coupler is shown. The coupler has a central coupling length of 8 mm and is characterised by launching TM-polarised light at $\lambda\approx 1550~$nm into one input port of the device. In the diagram the normalised transmission at the bar-state output is shown as function of the two voltages. It can clearly be seen that all possible splitting ratios can be achieved by a proper choice of the voltages.

For the modelling of the quantum circuits such electro-optically adjustable beam splitter are described by the usual beam splitter matrix Eq.~(\ref{BS}). Of course, in this case the two angles $\theta$  and $\xi$ depend on the applied voltages $U_1$ and $U_2$.

\subsubsection{Electro-optic polarisation converter}

For numerous quantum circuits polarisation converters, i.e.\ components which change the state of polarisation from TE to TM and vice versa,  are required. Several electro-optically adjustable polarisation converters in \LN\ have been demonstrated (see e.g.  \cite{Thaniyavarn},\cite{Alferness_PolConv,Heismann}). But most of these devices were either realised using X-cut, Y-propagating \LN\ \cite{Alferness_PolConv,Heismann} or Z-propagating \LN \cite{Thaniyavarn}. However, these two orientations are not compatible with a monolithic integration of nonlinear optical devices (like the PDC-sources), which require a Z-cut geometry for the periodic poling.  Such a periodic poling in Z-cut \LN\, can also be exploited to realise polarisation converters \cite{Huang_PolConv, Moeini_ECIO}. Its basic structure is sketched in \Fig{Active_Box}. It consists of a Ti-indiffused waveguide within a periodically-poled section with a poling period $\Lambda_{PC}$. Adjacent to both sides of the waveguides are electrodes deposited on the sample, i.e.\ an voltage $U$ applied to the electrodes leads to an electrical field mostly pointing along the crystalline Y-direction in the waveguide region, which gives rise to TE-TM coupling via the $r_{51}$ element of the electrooptic tensor. Please note, this conversion process requires phase-matching as well. The corresponding phase-matching condition is given by:
\begin{equation}
\Delta k_{PC}=\frac{2\pi}{\lambda}\left(n_{\mbox{\small TE}}(\lambda)-n_{\mbox{\small TM}}(\lambda)\right)-\frac{2\pi}{\Lambda_{PC}}
\end{equation}
with $n_{\mbox{\small TE/TM}}$ being the effective indices of the TE- and TM-mode, respectively. Similar to nonlinear optical interactions, this phase-matching requirement leads to a spectral dependence. A perfect conversion is achieved for $\Delta k_{PC}\approx 0$.

The efficiency of the conversion process is determined by the overlap between the optical fields and the electric field component along the crystalline Y-axis. As the field strength scales linear with the applied voltage, the coupling coefficient $\kappa$ is proportional to $U$. The TE-TM conversion process can easily be described using standard coupled mode theory.  Assuming that $A\TE(0)$ and $A\TM(0)$ are the amplitudes of the waveguide modes in TE and TM polarisation in front of the converter, these amplitudes evolve in the converter section according to see e.g.~\cite{Yariv_CMT}:
\begin{eqnarray}
\label{ConvEqn}\nonumber
A\TE(z)&=&\left\{A\TE(0)\left[\cos sz -i\frac{\Delta\beta}{2s}\sin sz\right]\right.\\ \nonumber
&&\left.-i\frac{\kappa}{s}A\TM(0)\sin sz\right\}\exp\left(i\frac{\Delta\beta z}{2}\right)\\ \nonumber
A\TM(z)&=&\left\{A\TM(0)\left[\cos sz +i\frac{\Delta\beta}{2s}\sin sz\right]\right.\\
&&\left.-i\frac{\kappa}{s}A\TE(0)\sin sz\right\}\exp\left(-i\frac{\Delta\beta z}{2}\right)
\end{eqnarray}
with
\begin{equation}
s=\sqrt{\kappa^2+\left(\frac{\Delta\beta}{2}\right)^2}.
\end{equation}

\begin{figure}[htbp]
\begin{center}
\includegraphics[width=15cm]{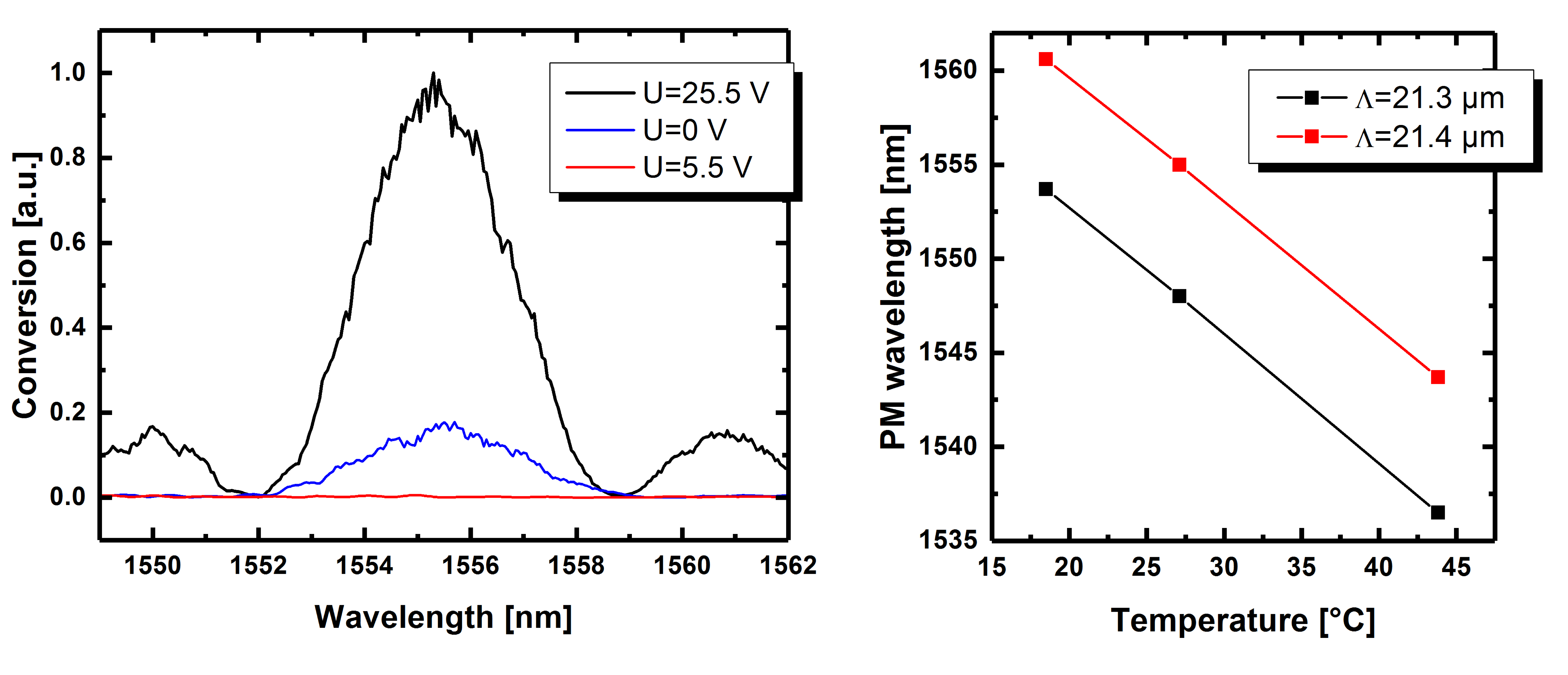}
\end{center}
 \caption{\label{PC_Spectra} Phase-matching and spectral characteristics of the electrooptic polarisation converter. Left: Measured spectral responses of a 7.6 mm long polarisation converter with $\lambda_{PC}=21.4~\mu$m at various drive voltages. Right: Measured phase-matching wavelength as function of the sample temperature for two different poling periods. 
}
\end{figure}

Such polarisation converters have been fabricated and experimentally characterised. In  the left diagram of \Fig{PC_Spectra} \ measured spectral characteristics are shown. The curves have been obtained using a 7.6 mm long structure with a poling period of $\Lambda=21.4~\mu$m. Spectra for different applied voltages are presented. Complete conversion is obtained with a driving voltage of $U=25.5~$V. The shape of the spectra shows the predicted sinc$^2$-behaviour with  $\Delta\lambda\approx3.2~$nm full width at half maximum (FWHM). Please note, that the spectral width is determined by the interaction length $L$. From Eq.\ (\ref{ConvEqn}) it can easily be derived that the FWHM is proportional to $1/L$. Thus, the spectral bandwidth as well as the matching of the central frequencies of all the elements on a circuit is a crucial design parameter, which has to be under full control in the fabrication processes and the experimental quantum optical setup. This becomes particularly obvious when considering the temperature dependence of the phase-matching as shown in the right diagram of \Fig{PC_Spectra}. The central phase-matching wavelength changes with temperature with a slope of $\approx -0.73~$nm/$^o$C. This strong temperature dependence might on one hand be used to fine tune the phase-matching to the desired operation point, but, on the other hand, means that homogeneity and temperature stability are of key importance, in particular, if several of such components are operated  simultaneously at different positions on the integrated circuit.

From the results shown in \Fig{PC_Spectra},
it is also remarkable that even for $U=0~$V some conversion can be observed. Obviously, this is due to some residual stress in the sample (which most probably arises during the periodic poling). Such stress can build up internal electric fields giving rise to the observed conversion.  At an applied voltage of $U\approx5.5~$V the conversion almost completely vanishes indicating that the internal fields are compensated by the field produced due to the applied voltage.

\begin{figure}[htbp]
\begin{center}
\includegraphics[width=7cm]{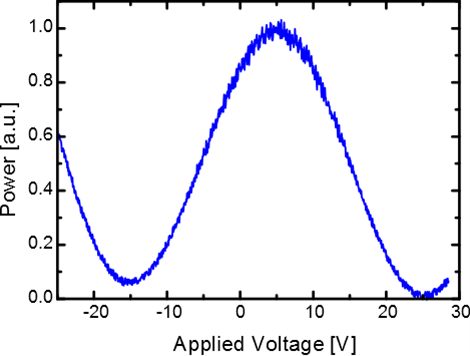}
\end{center}
 \caption{\label{PC_Power} Unconverted power vs applied voltage at phase matched wavelength. }
\end{figure}

At perfect phase-matching, the conversion efficiency can be tuned between 0 and 1 by adjusting the voltage. A corresponding measurement result is shown in \Fig{PC_Power}, where we varied the voltage and measured the unconverted power. As predicted from Eq.\ (\ref{ConvEqn}), we can clearly observed the sin$^2$-behaviour.

\begin{figure}[h]
\begin{center}
\includegraphics[width=7.5cm]{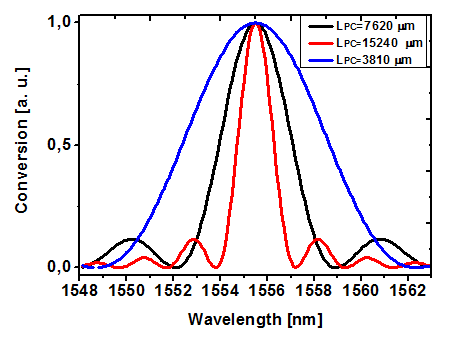}
\end{center}
 \caption{\label{Conversion} Calculated window of conversion for different PC lengths: 3810 $\mu$m (blue), 7620 $\mu$m (black) and 15240 $\mu$m (red). }
\end{figure}

For the modelling of the polarisation converter (PC) we describe the
action of the PC in the basis $\{a_{1H}^{\dagger}(\omega_{s,i}),a_{1V}^{\dagger}(\omega_{s,i}),a_{2H}^{\dagger}(\omega_{s,i}),a_{2V}^{\dagger}(\omega_{s,i})\}$  by a transformation matrix which depends on phasematching condition \cite{Heismann}
\begin{equation}
PC=\begin{pmatrix}
  \cos \phi+i a \sin \phi & -b\sin \phi & 0 & 0 \\
  b\sin \phi &  \cos \phi-i a \sin \phi & 0 & 0 \\
 0 & 0 & 1 & 0 \\
 0 & 0 & 0 & 1\\
 \end{pmatrix},
\label{PC}
\end{equation}
where $a=\frac{\frac{\Delta k_{PC}}{2}}{\sqrt{\kappa^2+(\frac{\Delta k_{PC}}{2})^2}}$, $b=\frac{\kappa}{\sqrt{\kappa^2+(\frac{\Delta k_{PC}}{2})^2}}$.  $\phi=\sqrt{\kappa^2+(\frac{\Delta k_{PC}}{2})^2} L_{PC}$ is the angle of polarisation conversion and  $L_{PC}$ is the length of the polarisation converter. For  $\kappa=\frac{\pi}{2 L_{PC}}$ and perfect phase matching  $\Delta k_{PC}=0$ one obtains a complete conversion, i.e.\  $\phi=\pi/2$.  In this case the polarisation is switched from horizontal to vertical and vice versa.
The PC matrix is the same for the signal and the idler photons.
It is important to note that since the PC depends on the phasematching condition, the corresponding
matrix is intrinsically frequency dependent and resulting from this frequency dependence, conversion happens only in a narrow wavelength range where $\Delta k_{PC}\approx 0$.
This means that the spectral shape of PDC photons must belong to this conversion region,
otherwise imperfect conversion efficiency will occur.
A window of conversion for a poling period of $\Lambda_{PC}\approx 21.4~\mu$m is defined by the
absolute value squared of the matrix element $|-b\sin \phi|^2$ 
and shown in \Fig{Conversion}.
With increasing PC section length the window of conversion gets narrower which may result
a degraded conversion efficiency.

\section{Advanced Circuit Example}

In previous section we reviewed and characterised a functional toolbox based on the  \LN\ platform. These functional components can now be used to construct complex integrated circuits for different specific quantum applications. To demonstrate the practicality of combinations and the interplay of all components on a single chip,
we present in this section a specific example, which can be considered as a benchmark system.
We show, how it is possible to design a circuit with a desired functionality and how the modelling tools
can be applied to predict the device performance.
In our example, we choose a circuit which includes a PDC-based photon-pair source and an asymmetric Mach-Zehnder like interferometer structure giving rise to HOM quantum interference.

\begin{figure}[htb]
\begin{center}
\includegraphics[width=8cm]{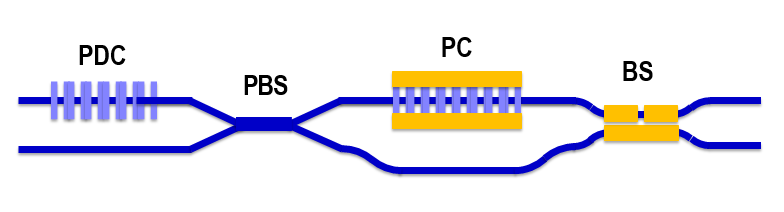}
\end{center}
\caption{\label{FullStructures} Schematical illustration of a possible implementation of the complete circuit. Proposed integrated quantum optical chip with monolithically integrated PDC-source, polarisation splitter (PBS), electrooptic polarisation controller (e/o-PC) and interferometer beam splitter (BS).}
\end{figure}

In \Fig{FullStructures} the structure of the monolithic quantum optical waveguide chip is sketched schematically. The current Ti-indifusion technique in \LN \ allows creating devices of 9-10 cm length and to put together up to 20 different elements. The number of cascade elements depends on length of each element and the actual design and functionality on-chip. For example, the typical BS with larger gap has about 1 cm length, the PBS with zero gap have lengths around 0.2-0.5 cm. The structure represented in \Fig{FullStructures} shows how different elements can be combined together. In a real fabricated chip the elements can be rearranged and the structure can be modified. Along with the above described optical elements this chip includes the free propagation sections which are described by free propagation matrices. Each free propagation matrix has the form 
\begin{equation}
FP=\begin{pmatrix}
  \exp[i \frac{\omega_s}{v_H} l_1] & 0 & 0  & 0 \\
   0 & \exp[i \frac{\omega_s}{v_V} l_1] & 0 & 0 \\
0 & 0 &  \exp[i \frac{\omega_s}{v_H} l_2] & 0 \\
 0 & 0& 0 &  \exp[i \frac{\omega_s}{v_V} l_2]\\
 \end{pmatrix},
\label{free}
\end{equation}
where $v_H=c/n_H(\omega_s)$ and  $v_V=c/n_V(\omega_s)$, $n_H$, $n_V$ are the refractive indices for horizontally- and vertically-polarised light,
$c$ is the speed of light in vacuum and $l_1$ and $l_2$ are the lengths of the free propagation in the upper channel $1$ and the lower channel $2$, respectively.
The same  free propagation matrix applies also to the idler photons when $\omega_s$ is
changed into $\omega_i$.

First, spectrally degenerate photon pairs  at telecom wavelengths are generated via type II phase-matched PDC. The orthogonally-polarised photons of the generated pairs are spatially separated by a PBS. By a polarisation converter the polarisation of one photon is rotated by 90$^o$.  Due to the different group velocities there is a  time delay between the photons in the various components of the chip, which can be compensated by different optical path lengths of the upper and lower branch in the subsequent part of the Mach-Zehnder interferometer.
Then, both photons of the pair are routed to the input port of a directional coupler acting as a BS. In this BS the two photons, which are now indistinguishable with respect to frequency, polarisation and time, interfere, which results in bunching  according to the HOM effect.

\subsection{Theoretical Description}

The total  transformation matrix for the setup sketched in \Fig{FullStructures} has the form
\begin{equation}
U_{total}=BS*FP_3*PC*FP_2*PBS*FP_1 ,
\label{transformation}
\end{equation}
where $FP_1$ corresponds to the free propagation between the end of the PDC section to the centre of the $PBS$ (lengths $l_1=l_2=y$),
$FP_2$ from the centre of the $PBS$ to the centre of the $PC$ (length $l_1=l$ for the upper channel and $l_2=L=l+\Delta l$ for the lower one),
and $FP_3$ from the centre of the $PC$ to the centre of the $BS$ (lengths $l_1=l_2=z$).
The free propagation from the BS to the detectors could also be included by considering an additional FP-matrix, however, since we consider in
the following the detection of two photons with identical polarisation this would just lead to an unimportant  overall phase factor if the propagation lengths to each of the two detectors are identical.
Due to the polarisation-dependent group velocities of the photons the compensation of the time delay can be performed by bringing an additional length to the second channel $L=l+\Delta l$.

The matrix $U_{total}$ describes the transformation between the input and the output states.
In the basis $\{a_{1H}^{\dagger},a_{1V}^{\dagger},a_{2H}^{\dagger},a_{2V}^{\dagger}\}$
the input state, see Eq.~(\ref{input}),
can be represented by
\begin{eqnarray}
\Ket{\psi_{PDC}}=\int d \omega_s d \omega_i F(\omega_s, \omega_i)
\nonumber\\
 U_0(\omega_s)
 \begin{pmatrix}
{a_{1H}^{\dagger} (\omega_s) } \\
 {a_{1V}^{\dagger} (\omega_s)} \\
{a_{2H}^{\dagger} (\omega_s) } \\
{a_{2V}^{\dagger} (\omega_s) } \\
\end{pmatrix}
\otimes
U_0(\omega_i)
 \begin{pmatrix}
{a_{1H}^{\dagger} (\omega_i)  }\\
 {a_{1V}^{\dagger} (\omega_i)} \\
{a_{2H}^{\dagger} (\omega_i)} \\
{a_{2V}^{\dagger} (\omega_i) } \\
\end{pmatrix} \Ket{0},
 \label{input1}
\end{eqnarray}
where 
\begin{equation}
U_0(\omega_s)=
\begin{pmatrix}
 1 & 0 & 0 & 0 \\
   0 & 0 & 0 & 0 \\
0 & 0 & 0 & 0 \\
0 & 0 & 0 & 0 \\
\end{pmatrix}
, \ \
U_0(\omega_i)=
\begin{pmatrix}
 0 & 0 & 0 & 0 \\
   0 & 1 & 0 & 0 \\
0 & 0 & 0 & 0 \\
0 & 0 & 0 & 0 \\
\end{pmatrix} ,
 \label{Us0Usi}
\end{equation}
are required to properly describe the photon pair generated in the type II PDC process.
 
The output state is defined by action of the total transformation matrix on the signal and idler subsystems.
However, the measuring procedure is performed on the output operators. Thus we have to use the reverse transformation
and express the initial operators in Eq.~(\ref{input1}) through the final ones, i.e.,
\begin{eqnarray}
\begin{pmatrix}
{a_{1H}^{\dagger} (\omega_s,i) } \\
 {a_{1V}^{\dagger} (\omega_s,i)} \\
{a_{2H}^{\dagger} (\omega_s,i) } \\
{a_{2V}^{\dagger} (\omega_s,i) } \\
\end{pmatrix}= U_{total}^{\dagger}(\omega_{s,i})
 \begin{pmatrix}
{f_{1H}^{\dagger} (\omega_s,i) } \\
 {f_{1V}^{\dagger} (\omega_s,i)} \\
{f_{2H}^{\dagger} (\omega_s,i) } \\
{f_{2V}^{\dagger} (\omega_s,i) } \\
\end{pmatrix} ,\ \ \ \ \ \
\label{inverse}
\end{eqnarray}
where $ f^{\dagger}$ and $f$ are the output creation and annihilation operators which describe the photons
after the action of the final BS, i.e., directly before reaching the detectors.
Substituting Eq.~(\ref{inverse}) in Eq.~(\ref{input1}) we can write the output state as
\begin{eqnarray}
\Ket{\psi_{out}}=\int d \omega_s d \omega_i F(\omega_s, \omega_i)
 U_0 (\omega_s)  U_{total}^{\dagger} (\omega_s)
 \begin{pmatrix}
{f_{1H}^{\dagger} (\omega_s) } \\
 {f_{1V}^{\dagger} (\omega_s)} \\
{f_{2H}^{\dagger} (\omega_s) } \\
{f_{2V}^{\dagger} (\omega_s) } \\
\end{pmatrix} \otimes
 \nonumber\\
 U_0 (\omega_i)  U_{total}^{\dagger} (\omega_i)
 \begin{pmatrix}
{f_{1H}^{\dagger} (\omega_i) } \\
 {f_{1V}^{\dagger} (\omega_i)} \\
{f_{2H}^{\dagger} (\omega_i) } \\
{f_{2V}^{\dagger} (\omega_i) } \\
\end{pmatrix} \Ket{0} . \ \ \ \ \  \ \ \
  \label{output}
\end{eqnarray}

The total transformation matrix for the signal frequency can be represented in the form 
\begin{eqnarray}
 U_{total}(\omega_s )= 
 \begin{pmatrix}
A_{11}^{HH}& A_{11}^{VH}& A_{21}^{HH}& A_{21}^{VH}& \\
\\
A_{11}^{HV}& A_{11}^{VV}& A_{21}^{HV}& A_{21}^{VV}& \\
\\
A_{12}^{HH}& A_{12}^{VH}& A_{22}^{HH}& A_{22}^{VH}& \\
\\
A_{12}^{HV}& A_{12}^{VV}& A_{22}^{HV}& A_{22}^{VV}& \\
 \end{pmatrix} ,
\label{Us}
\end{eqnarray}
where $A_{n n^{'}}^{\lambda \lambda^{'}}$ is the matrix element corresponding to the transition $n \lambda \rightarrow n^{'} \lambda^{'}$.
A similar matrix can be written for the idler frequency with coefficients denoted by $B_{n n^{'}}^{\lambda \lambda^{'}}$.
Altogether, the output state is given by
\begin{eqnarray}
\Ket{\psi_{out}} = \int d \omega_s d \omega_i F(\omega_s, \omega_i)\sum_{n,\mu}A_{1n}^{H \mu} (\omega_s)f^\dagger_{n,\mu}(\omega_s)
\nonumber\\
\times \sum_{n^{'},\mu^{'}}B_{1,n^{'}}^{V \mu^{'}} (\omega_i)f^\dagger_{n^{'},\mu^{'}}(\omega_i)\Ket{0},
\label{output}
\end{eqnarray}
where $n,n^{'}=\{1,2\}$ are the channel numbers and $\mu,\mu^{'}=\{H,V\}$ are the polarisation indices.
$A_{1n}^{H \mu} (\omega_s)$ and $B_{1,n^{'}}^{V \mu^{'}}  (\omega_i)$ are the coefficients which depend on the action of the optical elements $\phi, \theta, \xi, \alpha, \beta$ and on the lengths of the free propagations.
 
Measuring the coincidence probability at the detectors corresponds to calculating  the expectation value of simultaneous positive-operator valued measures (POVM) 
\begin{equation}
P_{b,c}^{\lambda, \lambda^{'}}= \Bra{\psi_{out}}M_{b}^{\lambda}\otimes M_{c}^{\lambda^{'}} \Ket{\psi_{out}},
\label{POVM}
\end{equation}
where the POVM operators are given by
\begin{eqnarray}
M_{b}^{\lambda}=\int d \omega_b d_{1,\lambda}^{\dagger} (\omega_b)\Ket{0} \Bra{0}d_{1,\lambda}(\omega_b) , \\
\nonumber\\
M_{c}^{\lambda^{'}}=\int d \omega_c d_{2,\lambda^{'}}^{\dagger} (\omega_c)\Ket{0} \Bra{0}d_{2,\lambda^{'}}(\omega_c) .
\label{POVMdef}
\end{eqnarray}
Here $d_{1,\lambda}^{\dagger}(\omega_b)$ and $d_{2,\lambda^{'}}^{\dagger}(\omega_c)$ are the creation operators of two detectors labeled as $b$ and $c$ and $\lambda$ and $\lambda^{'}$ are the polarisations which are measured by the detectors $b$ and $c$, respectively.
We assume that detector $b$ corresponds to the upper channel 1 and detector $c$ to the lower channel 2.
After a transformation the coincidence probability can be written in the simpler form
\begin{equation}
P_{b,c}^{\lambda, \lambda^{'}}=\int d \omega_b d \omega_c |\Bra{0} d_{1,\lambda}(\omega_b)d_{2,\lambda^{'}}(\omega_c) \Ket{\psi_{out}}|^2 .
\label{POVM1}
\end{equation}
Substituting the output state, Eq.~(\ref{output}), in Eq.~(\ref{POVM1})  and taking into account the commutation relation for the operators $[d_{n,\lambda}(\omega_b), f_{n^{'},\mu}^{\dagger}(\omega_s)]=\delta_{n,n^{'}}\delta_{\lambda,\mu}\delta(\omega_b-\omega_s)$ we obtain the following expression for the expectation value of simultaneous POVM
\begin{eqnarray}
P_{b,c}^{\lambda, \lambda^{'}}=\int d \omega_b d \omega_c |F(\omega_b, \omega_c)\times
\nonumber\\
\delta_{1,n} \delta_{\lambda, \mu} A_{1n}^{H \mu}(\omega_b)\delta_{2,n^{'}} \delta_{\lambda^{'}, \mu^{'}}B_{1,n^{'}}^{V \mu^{'}}(\omega_c)+
\nonumber\\
F(\omega_c, \omega_b) \delta_{1,n^{'}} \delta_{\lambda, \mu^{'}}B_{1,n^{'}}^{V \mu^{'}}(\omega_b)\delta_{2,n} \delta_{\lambda^{'}, \mu}A_{1n}^{H \mu}(\omega_c) |^2 .
\label{POVM2}
\end{eqnarray}

For the case of polarisation insensitive detectors, which we do not consider below,
the expectation value of the POVM corresponds to the sum of Eq.~(\ref{POVM1}) over the all polarisation combinations, i.e.,
 \begin{equation}
 P_{b,c}^{ins}=\sum_{\lambda, \lambda^{'}} P_{b,c}^{\lambda, \lambda^{'}}.
\label{insensitive}
\end{equation}

Below we evaluate the dependence of the coincidence probability on the length difference
of the upper and lower channels.
To vary this length difference experimentally one could use an external delay line, which would, however, mean that one leaves
the fully-integrated design.
Alternatively, one may implement a kind of discrete delay line also in the integrated design by adding several switchable spatially-separated PCs
in one channel and taking advantage of the different group velocities of horizontally- and vertically-polarised photons.

\subsection{Dependence of the Hong-Ou-Mandel Interference Parameters and Imperfections}

\begin{figure}[htb]
\begin{center}
\includegraphics[width=0.9\textwidth]{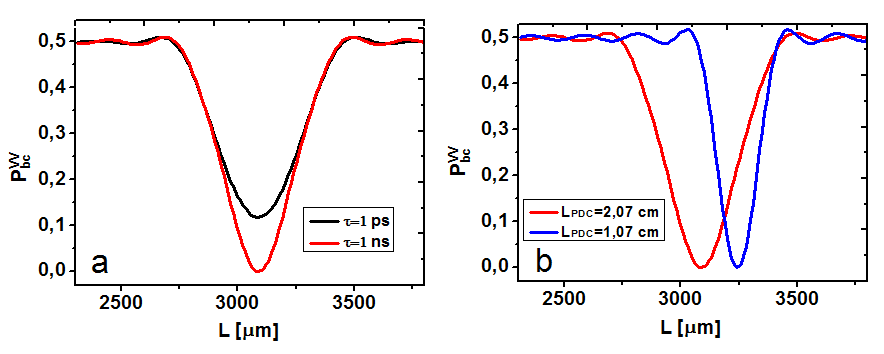}
\end{center}
\caption{The HOM dip as measured by the coincidence probability for detecting two vertically-polarised photons in the detectors for a perfect chip as sketched in \Fig{FullStructures}.
(a) Considering two pump pulse durations: $\tau= 1$ ns (red curve) and $\tau=1$ ps (black curve) for the same length of the PDC section $L_{PDC}=2.07$~cm, (b) Considering two PDC section lengths: $L_{PDC}=1.07$~cm (blue curve) and $L_{PDC}=2.07$~cm (red curve) for the same pulse duration ($\tau=1$ ns)} \label{fig:HOMTPA}
\end{figure}

For the case of an ideal integrated device as sketched in \Fig{FullStructures}, the generated two photons become fully indistinguishable when they reach the BS. Modifying the time delay between the photons, by, e.g.\ changing the length of the lower channel, changes  the coincidence probability of the photons measured in the detectors. If this time delay equals zero, the two photons are fully indistinguishable and the coincidence probability drops down close to zero, see Fig.~\ref{fig:HOMTPA}. By increasing the time delay the photons become partially distinguishable and the coincidence probability increases and reaches $0.5$ in the limit of a large time delay when the photons are totally distinguishable. 

A typical profile of the HOM dip is depicted in Fig.~\ref{fig:HOMTPA}. This figure presents the expectation value of the POVM with $\lambda, \lambda^{'}$ both equal to $V$ corresponding to two vertically-polarised photons and fully ideal optical elements. The HOM profile depends on the structure of the TPA function.
Fig.~\ref{fig:HOMTPA}(a) visualises the dependence of the HOM dip on the duration of the pump pulse.
In general, the TPA given by Eq.~(\ref{TPA}) is not fully symmetrical with respect to the exchange of the signal and idler variables due of the different refractive indices for the horizontally- and vertically-polarised light. If, however, the pulse duration is sufficiently long the TPA becomes very narrow in the $\omega_s=\omega_i$ direction and in this case the TPA becomes very close to a symmetrical function. This means that the minimum of the HOM dip will be very close to zero. For shorter pulse duration the non-symmetrical character of the TPA becomes more evident which leads to an increased value of the minimum of the coincidence probability.

With increasing  length of the PDC section, see Fig. ~\ref{fig:HOMTPA}(b), the TPA becomes narrower in the frequency domain and consequently the HOM dip becomes broader in the time domain. Simultaneously the position of the HOM dip is shifted in length due to the increasing time delay between the photons accumulated already in the PDC section. This dependence on the length of the PDC section arises since the photon pairs are predominantly generated in the centre of the PDC section as modeled by the imaginary exponent in Eq.~(\ref{TPA}).

\begin{figure}[htb]
\begin{center}
\includegraphics[width=0.5\textwidth]{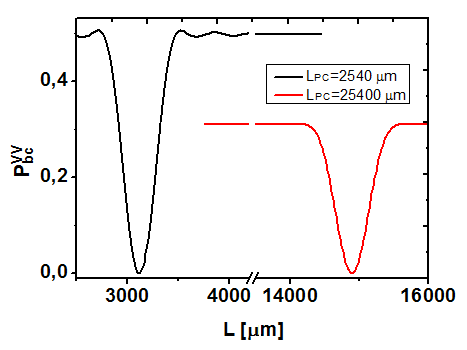}
\end{center}
\caption{The frequency dependence of the HOM dip is connected with the length of the PC. The HOM dip for  $L_{PC}=2540 ~\mu$m (black curve) and $L_{PC}=25400 ~\mu$m (red curve). With increasing length of the PC the window of conversion gets narrow. That leads to worse conversion of side frequencies in the TPA,the HOM dip becomes broader, the maximum value decreases.  } \label{fig:HOM_vs_PC_length}
\end{figure}

\begin{figure}[htb]
\begin{center}
\includegraphics[width=1.0\textwidth]{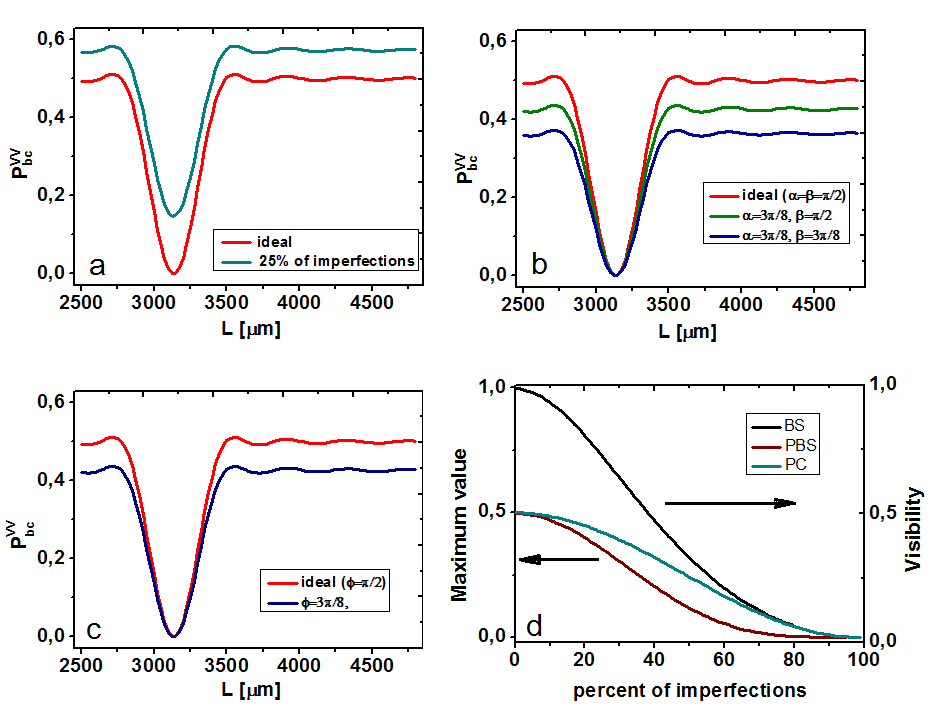}
\end{center}
\caption{The influence of imperfections of different elements on the coincidence probability.
In (a)-(c) the red curves corresponds to the ideal case, whereas in the other curves 25$\%$ of imperfections in (a) the BS, (b) the PBS (here the green curve corresponds to the 25$\%$ of imperfections for only one polarisation whereas the blue curve is calculated for 25$\%$ of imperfections for both polarisations), (c) the $PC$.
(d) The visibility as function of the BS imperfections (black curve)
and the maximum values of the coincidence probability as function of the PBS (red curve) and the $PC$ (cayan curve) imperfections.} \label{fig:imperfections1}
\end{figure}

In a real device the considered optical elements will typically not work perfectly but their operation characteristics might deviate from
the ideal case due to the limited accuracy of the fabrication process. 
In addition, also a certain degree of losses arising, e.g., from bending the waveguides or from the contacts that are required to apply the voltage,
will be present. Such type of imperfections are statical imperfections because they do not modified with changing external conditions. But simultaneously there are present dynamical imperfections, for example, dependence the HOM interference visibility on temperature, that requires sufficiently higher level of accuracy in the experiment. In our work, the waveguide losses are less than 0.1 dB/cm and can be neglected in comparison with the coupling losses among different bulk sections.
It is very important to know the limit of imperfections of the different optical elements until which quantum effects in the coincidence probability are still observable which corresponds to a minimum of the HOM dip smaller than the classical limit of $0.25$.

With respect to imperfections and losses the most sensitive considered optical element is the PC.
The PC has a window of conversion which corresponds to a principal limitation of the conversion efficiency for a fixed TPA profile. As usual the length of the PC is chosen such that the  window of conversion is significantly broader than the spectral width of the TPA providing a good conversion, see black curve in
Fig.~\ref{fig:HOM_vs_PC_length}.
However, with increasing the length of the PC the window of conversion gets narrower which leads to non-perfect conversion of side frequencies and broadening of the HOM dip;
simultaneously, the maximal value of coincidence probability decreases, see red curve in
Fig.~\ref{fig:HOM_vs_PC_length}.

The modifications of the HOM curves due to imperfections are different for each of the optical elements. 
The profiles of the HOM dip with $25\%$ imperfections are presented for the case of the
BS in Fig. ~\ref{fig:imperfections1}(a),
the PBS in Fig. ~\ref{fig:imperfections1}(b)
and the $PC$ in Fig. ~\ref{fig:imperfections1}(c), respectively.
In Fig. ~\ref{fig:imperfections1}(a)-(c) the red curves correspond
to the ideal case when all optical elements work perfectly
and in each figure we include imperfections of only a single optical element and assume that the other ones work perfectly.
25$\%$ of imperfections is defined in relation to the entire range of the respective angle which describes the element's operation.
 For example, the range of the BS is $\theta =[0,\pi/4]$, where $\theta =\pi/4$ corresponds to the perfect case of a 50/50 beam splitter
and $\theta =0$ to the totally imperfect case when light is just transmitted by the BS without any reflection.
Therefore, 25$\%$ of imperfections means that $\theta =3\pi/16$.
Analogously, for the PBS and PC the range of definitions are $\phi =[0,\pi/2]$ and $\alpha, \beta =[0,\pi/2]$ and therefore 25$\%$ imperfections corresponds to $\phi=\alpha=\beta=3 \pi/8$ (in the case of PC this value of imperfections corresponds to the perfect phase matching condition $\Delta k_{PC}=0$).

Fig.~\ref{fig:imperfections1}(a) shows that with imperfections of the BS the coincidence probability goes up and thus the HOM dip is weakened.
In the totally imperfect case, i.e., when the BS has 100$\%$ transmission and 0$\%$ reflection, the coincidence probability goes to 1.
Fig.~\ref{fig:imperfections1}(b) and (c) present the effects of imperfections in the PBS and in the PC.
In both cases imperfections reduce the coincidence probability and in the limit of a totally imperfect PBS or PC it will go to zero.
This is due to the fact that with increasing imperfections the PBS starts to separate photons with different polarisation not perfectly which leads to a mixture of photons with different polarisation in both channels that destroys the two-photon quantum interference.
Increasing imperfections of the PC correspond to a non-perfect conversion of the polarisation in upper channel which leads to a reduced indistinguishability and thus a weakening of the two-photon quantum interference.
Fig. ~\ref{fig:imperfections1}(d) displays the change of the visibility as a function of the imperfection of the BS and
the change of the maximum value in the coincidence probability as function of the imperfections of the PBS and the PC.
In agreement with the above arguments, 
the visibility of the HOM dip 
\begin{equation}
\nu = \frac{P_{b,c}^{\lambda, \lambda^{'}}(\infty)-P_{b,c}^{\lambda, \lambda^{'}}(0)}{P_{b,c}^{\lambda, \lambda^{'}}(\infty)}
\label{visibility}
\end{equation}
decreases to zero for a totally imperfect BS, whereas the maximum coincidence probabilities decrease from the ideal value of $0.5$ towards
zero with increasing imperfections of the PBS and the PC. From Fig.~\ref{fig:imperfections1} it follows that the quantum interference is observed even for 25 $ \% $ of imperfections in optical elements. The recently achieved fabrication technique allows holding losses in a few percent range (2-3 $\%$) that makes expect a full experimental realisation of an on-chip interferometer in the nearest future.

\begin{figure}[htb]
\begin{center}
\includegraphics[width=1.0\textwidth]{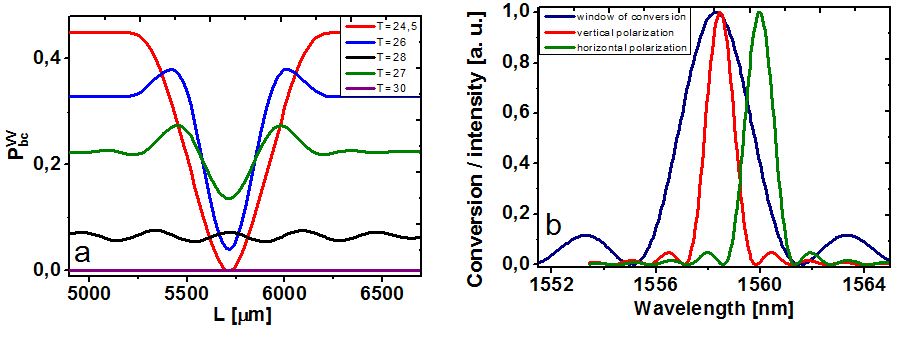}
\end{center}
\caption{(a)The temperature dependence of the HOM dip. (b) The window of conversion of the PC (blue curve), the normalised intensity profiles of vertically- (red curve) and horizontally-polarised (green curve) photons for a fixed temperature $T=26$, length of PC $7620 \mu m$} \label{fig:temper}
\end{figure}

Except for static imperfections, a crucial role in HOM visibility is played by dynamical imperfections, for example, the temperature. The temperature dependence of the HOM dip is presented in  Fig.~\ref{fig:temper} (a). The red curve corresponds to the ideal situation, when the perfect phasematching occurs for the degenerate wavelengths (for the considered poling period it is observed for the temperature $T=24.5$). With increasing temperature the refractive indices of the pump, signal and idler photons increase which leads to the changed phasematching  for a fixed poling period and pump wavelength. In turn, the  non-zero phasematching condition in the PDC section corresponds to the realisation of non-degenerate generation of the signal (horizontally-polarised) and idler (vertically-polarised) photons, see green and red curves in Fig.~ \ref{fig:temper} (b). Simultaneously the non-zero phasematching leads to a shifting of conversion window of the PC. As a result, with increasing the temperature the conversion of the horizontally-polarised photon becomes worse, see Fig.~\ref{fig:temper} (b), green curve, which leads to reduction of the visibility of the HOM dip. For a temperature far away from the ideal situation the conversion of the horizontally-polarised photon does not occur anymore and the HOM dip is vanishes. That is why the HOM interference is strongly temperature dependent and temperature stabilisation is an important point for the experimental realisation the on-chip HOM interference and the overall  design of circuit with multiple elements.

\section{Conclusions}
We have discussed advanced integrated circuits realised in \LN\ for quantum optical applications. With a few basic components one can realise a versatile toolbox which can be used to design a variety of quantum circuits with different functionalities. As an example, we propose a scheme of an integrated quantum interferometer  based on the \LN\  platform. 
The theoretical description of such an interferometer is based on unitary transformations for each component.
The influence of the TPA profile and of imperfections of each of the individual components on the final two-photon interference has been studied.
Quantum interference effects are predicted for a realistic experimental parameters.
By this work we provide guidelines and modelling tools which should be helpful to pave the way towards the implementation of  advanced integrated quantum circuits harnessing the potential of \LN.

\section*{Acknowledgement}
This work was funded by the Deutsche Forschungsgemeinschaft (DFG)
via TRR~142/1, project C02.
P.~R.~Sh. thanks the state of North Rhine-Westphalia for support by the
{\it Landesprogramm f{\"u}r geschlechtergerechte Hochschulen}.

\clearpage
\end{document}